\documentclass[12pt,aps,preprint,showpacs,superscriptaddress]{revtex4}  
\usepackage[draft]{hyperref} 
\usepackage{graphicx}
\usepackage{setspace}
\usepackage{amsmath}
\usepackage{amssymb}
\usepackage[english]{babel}

\begin{document}

\title{
Detection of short DNA sequences with DNA nanopores
}

\author{Luyan Yang, Christophe Cullin and Juan Elezgaray}

\affiliation{ Centre Paul Pascal, UMR 5130, CNRS,  Av. Schweitzer, 33600 Pessac, France }

\begin{abstract}
Several
studies suggest strong correlation between different types of cancer and the relative concentration of short
circulating RNA sequences (miRNA). Because of short length and low concentration, miRNA detection is
not easy. Standard methods such as RT-PCR require both the
standard PCR amplification step and a preliminary additional step of reverse transcription. In this paper, we investigate
the use of DNA nanopores as a tool to detect short oligonucleotide sequences at the single molecule level.
These nanostructures show two different conformations depending on the presence of  DNA analogues of miRNA sequences.
By monitoring current across a lipid bilayer, we show that this change of conformation translates to different
levels of conductivity.
\end{abstract}

\maketitle

\section{Introduction}
Micro-RNAs (miRNA) were first discovered in the nematode C.elegans \cite{ref2}, and subsequently found in
practically all eukaryotes. Despite their small size (length of the sequence between 19 and 24 nucleotides),
these single strand, non-coding RNAs play a role in the regulation of the genetic expression, through their
capacity to hybridize with the 3’UTR of specific target mRNA (messenger RNA). Therefore, the specific
function of each miRNA is strongly sequence-dependent, and the search of the associated targets is a non
trivial problem. As shown in a recent review \cite{ref3}, miRNAs are known to be associated with the normal
development and function of the organism,  but are also involved in diseases. Several lines of evidence highlight the putative implication
of miRNA in the physiopathology of numerous cancers. This includes the miRNA dysregulations observed in
tumors (over- or under-expression), that led to the concept of “oncomiR” \cite{ref4}, and to the use of miRNA for the
classification of tumors origins \cite{ref5}. Other arguments rely on the fact that widely expressed oncoproteins (e.g.
MYC) or tumor suppressors (e.g. P53) regulate the transcription of specific miRNA that, in turn, modify the
translation of cancer-associated genes \cite{ref4}. More recent findings revealed the existence of circulating miRNA,
especially in tumor-bearing animals or patients. It is likely that they derive from the dying cells of the tumor,
or are actively excreted in exosomes \cite{ref6}.

The attractiveness of extracellular miRNA as cancer biomarkers relies on their stability and their
dysregulation in the diseased cells. While bound to Argonaute proteins, miRNAs are stable in the
extracellular environment after release from cells, whether as unprotected ribonucleoprotein complexes or
within membranous vesicles. The released miRNA can be detected and quantitated as part of the
“miRNome” of a biological fluid, such as plasma or serum. If the miRNA expression in a tumor are reflected
in the circulation, invasive tissue biopsies could be replaced by straightforward assays of easily obtained
blood products, and early warnings of tumorigenesis might be possible.
Because of their short sequence and low concentration, miRNA detection is intrinsically difficult. Methods for this measurement can
be classified in three classes:
\begin{itemize}
\item[1-] PCR based methods requiring a preliminary reverse transcription step, which sets the initial
minimal total miRNA amount required to ~50ng.
\item[2-] PCR free methods, for instance based on electrochemistry or optical (SPR) detection, do not
require the reverse transcription step, and present limits of detection in the fg ($10^{-15}$ g) range. Similar
sensitivity can be reached with optical, single molecule methods, such as fluorescence correlation
spectroscopy (FSC) \cite{ref7}. In turn, all these methods require very sophisticated instrumentation. For this
reason, none of them has emerged as a practical alternative to more standard, PCR based methods.
\item[3-] Next-generation sequencing-based methods, that allow the detection of every miRNA species
expressed in the tissue or cells of interest, but also necessitate around 10-100ng of total miRNA.
\end{itemize}

This work is based on a completely different  approach, which in other contexts has been called
'stochastic detection' \cite{stox}. In this approach, two liquid media are separated by a membrane (lipid bilayer) in which
channels such as transmembrane proteins can insert. Stochastic detection is based on the correlation
between the presence of some analyte (here, miRNA) and the current across a single channel. A good
example of stochastic detection is the method developed  \cite{ref1} for
DNA  sequencing: a single stranded DNA going through a single nanopore channel modulates the
current through it and can be correlated with the DNA sequence.

In the past few years, DNA-based nanostructures \cite{nano1,nano2,nano3} have been developed that mimic naturally
occurring membrane proteins \cite{nano4,nano5,nano6}. These nanostructures can 
 also interact with lipid membranes. As compared to protein channels, they can be easily modified in terms of geometry or functionalization. 
In this report, we  adapt a recently published DNA
construction \cite{pore1}, the conductivity of which can be modulated in the presence of specific
oligonucleotides (DNA or RNA). The method therefore falls into the category of single molecule, PCR free
methods. 

The nanopore structure  in \cite{pore1} is formed by 15 DNA strands which fold into six intertwined helices 72nt long.
 Some of the
DNA strands that form the nanopore are covalently linked to cholesterol moieties to allow insertion into  lipid bilayers. 
 The design of ref.\cite{pore1}
includes a mechanism to regulate the current across the nanopore. This mechanism is based on a single
stranded DNA (hereafter called sentinel), linking two helices of the nanopore, which can be in two possible
states. In the absence of an input signal (the complementary sequence of the sentinel), this strand is in a globular, floppy
state: this is the closed state. When the input signal is present, it hybridizes to the sentinel, increasing its mechanical tension
and pushing apart two helices. The nanopore is in an open state with a larger inner diameter and electric conductivity.
 The conductivity difference between closed and open states depends on the geometry of the nanopore. In ref.\cite{pore1} we showed
that a 30nt long input sequence induces a measurable conductivity change. Unfortunately, this is not the case for a 22nt long sequence, as the
ones we are targeting here. This can be easily understood: a double
stranded DNA of 22 base pairs forms a stiff double helix 7nm long. This is intended to exert a
mechanical tension upon two helices that are separated by $\sim 6$nm. The present sensitivity of the method
precludes the detection of such small effect. 

Our goal here is to be able to detect single miRNA strands such as miR-21, a 22 nucleotide long miRNA involved in cancer
(thyroid, breast and colorectal) \cite{ref9}. To facilitate sample preparation and handling, we will use the DNA analog of miR-21, 
with sequence: TAGCTTATCAGACTGATGTTGA.
In the following, we describe a modification of the initial  structure \cite{pore1} where the regulation of conductivity
is made by changing the effective length of the nanopore. We successively describe nanopore structure and synthesis,
a method to form stable lipid bilayers on which nanopores can insert, finally a characterization of electrical signature as a function
of the presence of short oligonucletides.

\section{Methods and results: nanopore structure}

\begin{figure}
\includegraphics[scale=0.8]{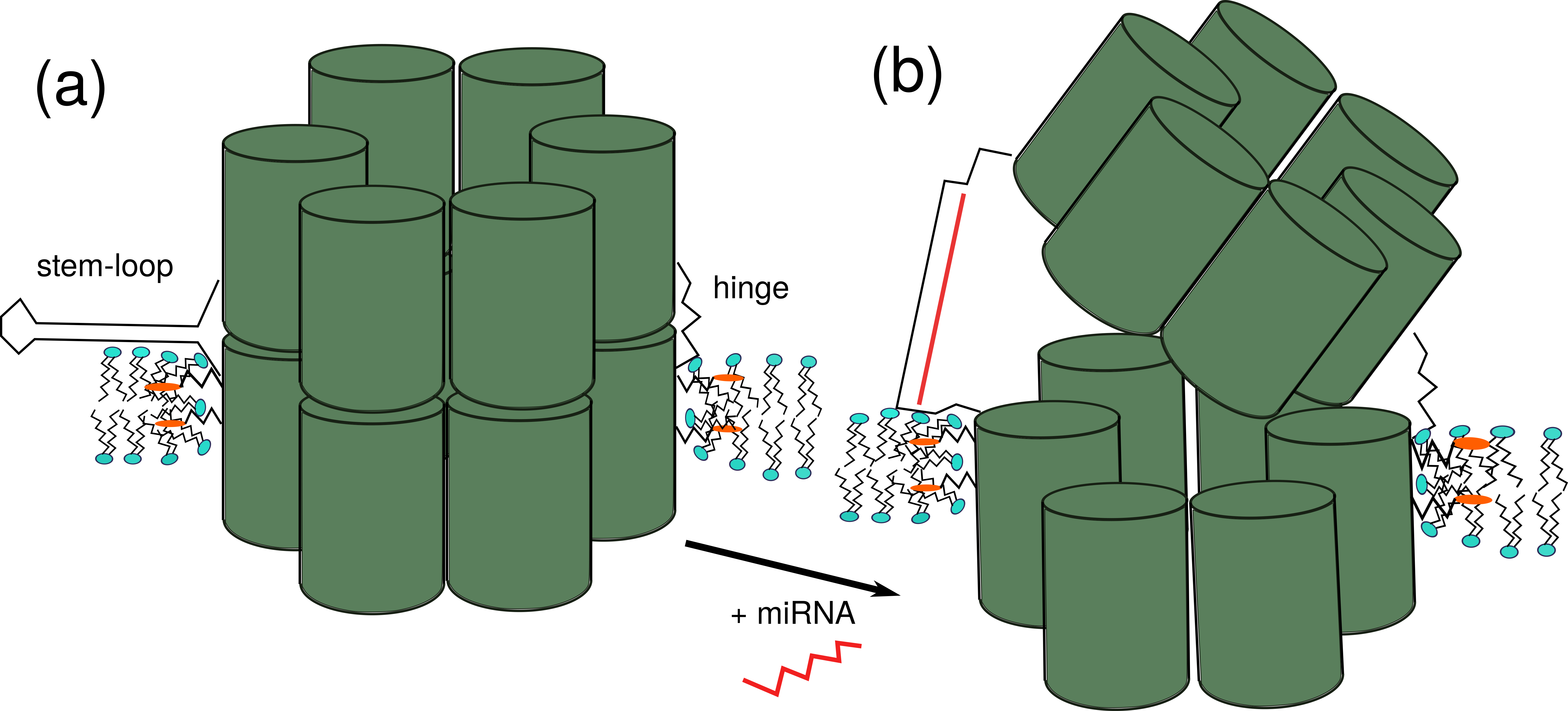}
\caption{Schematic representation of the opening mechanism: each cylinder represents a double helix. DNA nanopore is inserted
into a lipid bilayer thanks to cholesterol modifications (orange ellipses). (a) Closed state: the stem loop imposes a short distance
between two of the helices. (b) Open state: upon addition of miR-21, the stem loop unfolds giving rise to a mixed single and double stranded
linker which pushes the two halves apart.}
\label{fig_schema}
\end{figure}

The basic design of the nanopore is inspired from that in ref. \cite{How13}. In short, six double helices
are linked to form a barrel with hexagonal cross section. Strands thread between  double
helices, forming Holliday-type crossovers which increase structural stability. Our goal is to obtain a nanotube composed of two halves
linked by a hinge and a locking mechanism
(cf Fig. \ref{fig_schema}). To design the latter, we take inspiration from ref. \cite{box}. In this work, the goal was to trigger the opening and closing of a 3D origami box. The authors
used two locks, each including a stem-loop structure with a 8nt  loop. Upon addition of the opening key, an
oligonucleotide complementary to a subset of the lock, the lock opened. We will use a similar mechanism here, as illustrated in Figure. \ref{fig_schema}.

\begin{figure}
\includegraphics[scale=0.8]{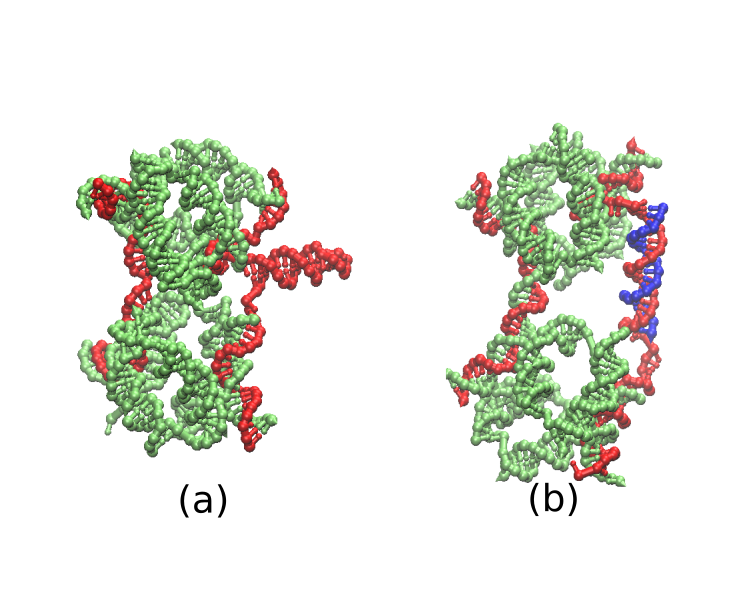}
\caption{Schematic representation of the opening mechanism as modeled with oxDNA software. In this illutration, each
nucleotide is represented by two sites, one centered on the phosphate group, the other centered on the nucleobase. 
Strands that form the hinge or the stem-loop locking
mechanism are in red. The input signal is in blue. (a) Nanopore in the absence of input signal. (b) Nanopore
hybridized to the input signal. }
\label{fig_oxdna}
\end{figure}

The nanopore is composed of two barrels. Each barrel is formed of six double helices arranged around a 2nm lumen with hexagonal
cross-section. The staple design contains Holliday junction crossovers to increase the stability of the ensemble. The
two barrels are linked by a hinge and a locking mechanism including a stem-loop structure.
 The stem is a 21nt double
helix, the loop is 10nt long. In the closed state, the stem-loop  effectively imposes a short distance ( $\sim 2$nm) between two  of the  helices. 
The transition to the open state requires an input signal which can
bind to the exposed nucleotides of the loop and open the stem through a strand displacement
process. The stem loop becomes a long single stranded loop with a 21nt long double stranded section. This   acts as an
entropic spring, effectively increasing the distance between the two previously close helices. The length of the loop was  optimized
by computing (using Nupack \cite{nupack}) the percentage of open structures.  In Figure \ref{fig_oxdna} are shown the results 
of two oxDNA \cite{oxDNA} simulations, where
the nanopore was simulated  in closed and open states. Nanopore was assumed to be in solution, the cholesteryl modifications or any interaction with
lipids were not taken
into account. As shown in Fig. \ref{fig_oxdna}, thermal fluctuations push the nanopore configuration far from the ideal  
 hexagonal arrangement. Still,
 these simulations give support to the idea that the opening of the stem loop 
can significantly perturb the geometry of a nanopore even for short input signals such as mi2R-21. Also, the simulations suggest
that the  stem loop structure when attached to the nanopore is reasonably stable against thermal fluctuations. 

\begin{figure}
\includegraphics[scale=0.8]{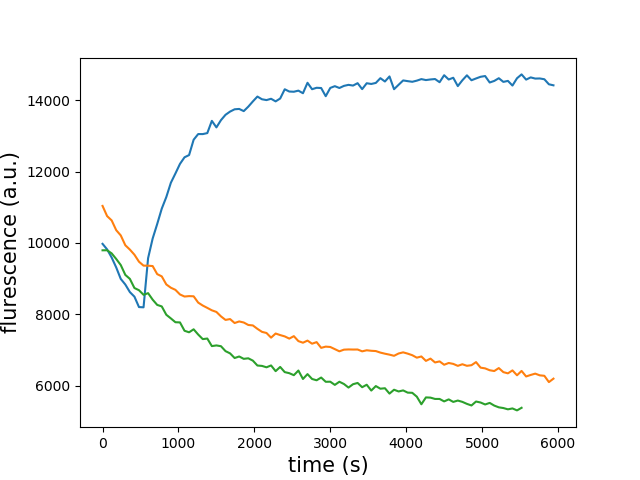}
\caption{Fluorescence (660nm) recorded as a function of time for three different samples, all of them containing 100nM solution of
nanopore, modified with a Cy5-BHQ2 couple. Before t=600s, each sample only contains the nanopore. At t=600s, an input signal
was added. Blue: addition of miR-21. Orange and green: addition of two random sequences, 32nt long.  }
\label{fig_fluo}
\end{figure}

To ascertain the opening mechanism and provide a proof of the detectability of the associated shape modification, we first used a fluorophore - quencher to monitor the FRET efficiency in the presence of the signal nucleotide. Two of the nanopore strands were modified with respectively Cy5 and BHQ2.
In the absence of input signal, the nanopore should be in the closed state, in which BHQ2 quenches the fluorescence emission of Cy5. Adding
miR-21 increases drastically the fluorescence recorded at 660nm, as can be seen in Fig. \ref{fig_fluo}. The same illustration shows that the
addition of two random sequences, 32nt long, had no effect on the nanopore opening.

\begin{figure}
\includegraphics[scale=0.8]{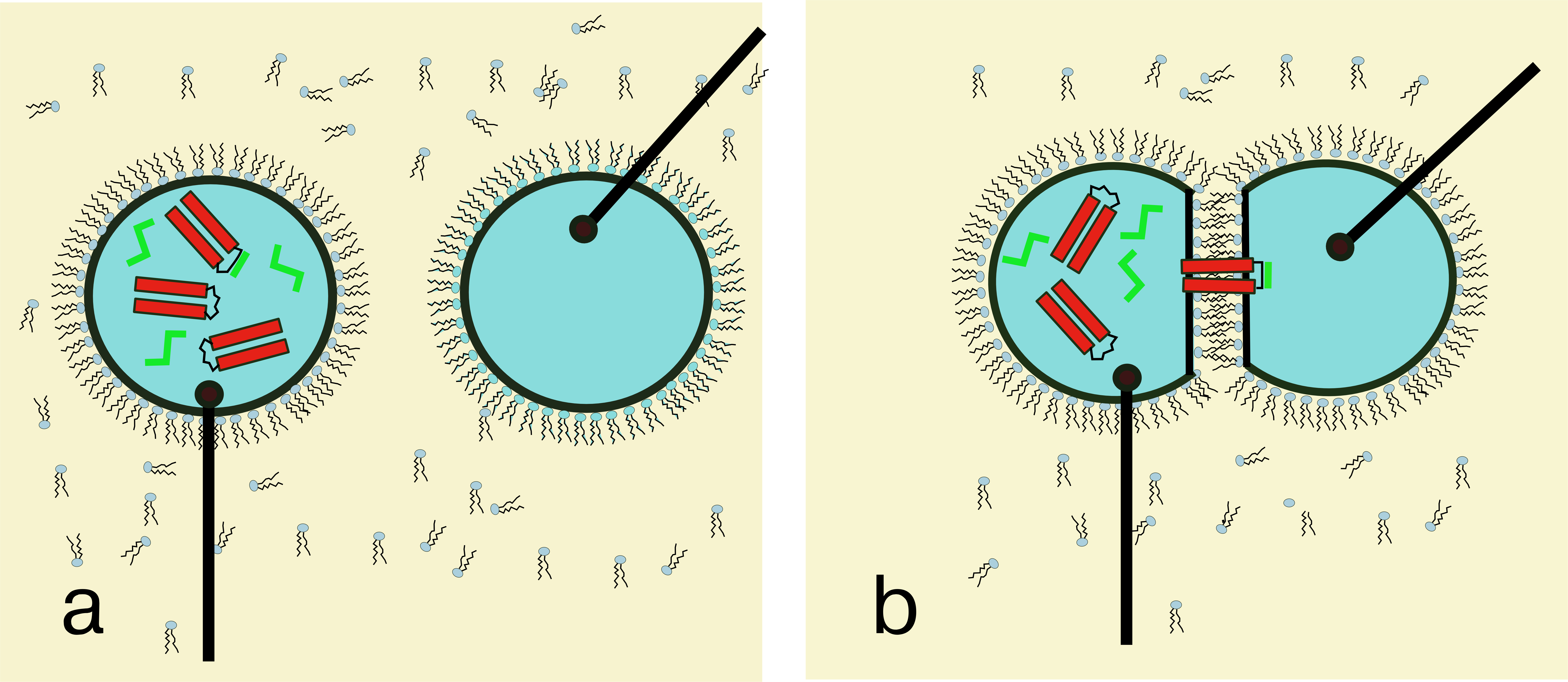}
\caption{Illustration of the droplet interface bilayer (DIB) method. Two aqueous droplets are immersed in an oil bath (yellow)
containing lipids in solution. Droplet position is monitored through electrodes (black segments). (a) Before contact, a monolayer
forms at the surface of each droplet. After contact (b) a bilayer forms. Nanopores are contained in only one of these droplets (red rectangles).
}\label{fig_dib}
\end{figure}

DNA nanopore structures can interact with lipid bilayers when modified with hydrophobic  moities. In the pioneering work of Simmel and coll. \cite{Simm12},
the authors showed by TEM imaging how a large origami structure, modified with cholesteryl (cholesterol is attached to the deoxyribose via a
six carbon spacer), was able to insert into a lipid vesicle. Subsequently,
several teams also showed how similar structures could interact with locally planar bilayers by recording  the current across bilayers.
The interest of this configuration is the possibility to detect the insertion of single nanopores. For the planar configuration 
two main options can be distinguished. The formation of a black lipid layer has been used in refs. \cite{How13, black}. In this configuration, two compartiments are separated
by a hydrophobic wall with a tiny hole. Painting lipids around the hole, then hydrating the system leads to the formation of a lipid bilayer.
Alternatively, the so-called droplet interface bilayer (DIB) \cite{bayley} configuration (Figure \ref{fig_dib}) deals with two aqueous droplets immersed in an oil bath containing lipids in solution.
When the two droplets are not in contact, a lipid monolayer forms around each droplet. The position of each droplet can be monitored through electrodes
connected to micromanipulators. If the two droplets are brought into contact, a lipid bilayer  quickly forms at the intersection. 
As previously noticed \cite{Dibcapa}, the stability
of this interface is remarkable,  although it strongly depends on lipid and oil composition. We used previously \cite{pore1}
a 'patch-clamp' approach to obtain a bilayer, patching small pieces of giant unilamellar vesicles with a micropipette. As compared to this latter method,
the use of DIBs is a much more robust approach with the disadvantage that not all lipid compositions can be explored. Further details are given in the
SI.

We considered two approaches to enhance nanopore insertion into bilayer. 
Four strands were elongated with a common sequence, 15nt long, to which a complementary
oligonucleotide modified with cholesteryl could bind. Experimentally, the insertion frequency of these structures into DIBs was very low. The second, more
successful method, enhanced the insertion by adding two biotin modifications in one side (cf. Fig. \ref{fig_schema}) of the nanopore in addition
to the four cholesteryl modifications. In this second strategy, DIB system
was asymetric. One of the droplets, connected to ground,  contained the nanopore. The other droplet, connected to the prove electrode, contained
a solution of streptavidin. The biotin-streptavidin interaction is a classical biological tool to bind two partners. Our intuition was that
in the event of a nanopore insertion, biotins would bind to streptavidin, thus  maintaining nanopores in close proximity of the bilayer. 
We also hypothetized that
the eventual transport of streptavidin across the DIB bilayer would be negligible.

\begin{figure}
\includegraphics[scale=0.8]{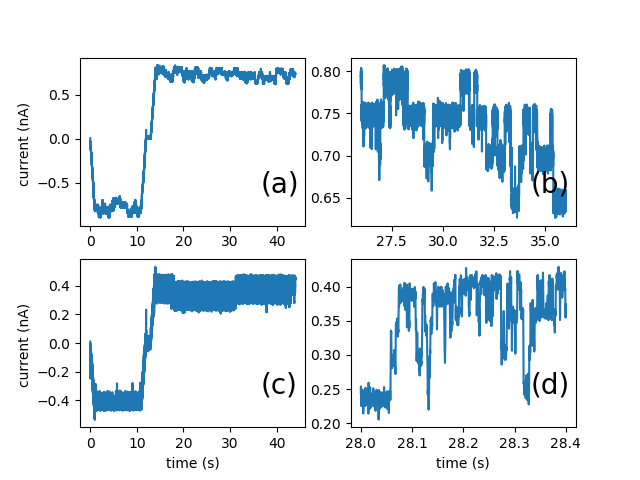}
\caption{ Time recording for a typical DIB experiment where voltage is varied between -30mV and 30mV.  (a)(b) Slow dynamics regime (c)(d)  Fast dynamics regime. (b) and (d) are zoomed images of (a) and (c), respectively.}
\label{fig_record}
\end{figure}

A typical experiment started by the insertion of agarose coated electrodes on each of the droplets. Then, the droplets were brought into
 contact by monitoring
 electrode position. A lipid bilayer  has a well defined capacitive response to a short (10ms) 10mV pulse. Its formation
could be easily monitored, usually it took less than one
minute after droplet contact. 
After stabilization of the bilayer's resistance, we imposed alternatively positive (30mV) and negative (-30mV) potentials between ground
and control electrode with very slow frequency (60s). During the positive potential phase, nanopores were expected to be driven towards the bilayer. 
Correspondingly, a negative potential would tend to remove them. Figure \ref{fig_record} illustrates two typical situations we encountered. 
Current time recordings showed step-like profiles with an essentially stable baseline. We observed several  characteristic time intervals
 between jumps, with no evident link with experimental conditions. Fast transitions (jump frequency around 100Hz) were much more frequent than
the slower transitions (jump frequency around 10Hz) displayed in panels \ref{fig_record}(a) and \ref{fig_record}(b).
The
average succes ratio, defined as the number of current recordings where jumps could be observed divided by the total number
of current recordings, was rather low (less than 10\%). Unsuccessful recordings gave usually a flat signal (no jumps), or the interface was 
unstable leading to data difficult to interpret (with a vast majority of flat signals). In the absence of nanopores, no jumps were observed at all.

To count the number and measure conductivity jumps, we used a Hidden Markov Model (HMM) \cite{hmm}. 
Given a time recording,
HMM approximates it by a sequence of $N_s$ states, the values of which are optimized to minimize the difference between the sequence and the given
time series. The number $N_s$ is a free parameter of the model. As shown in Fig. \ref{fig_record}, 
HNN approximation  seems to be well adapted to the current recordings as a well defined base line
can be easily found and the number $N_s$ appears to be well defined. 

\begin{figure}
\includegraphics[scale=0.8]{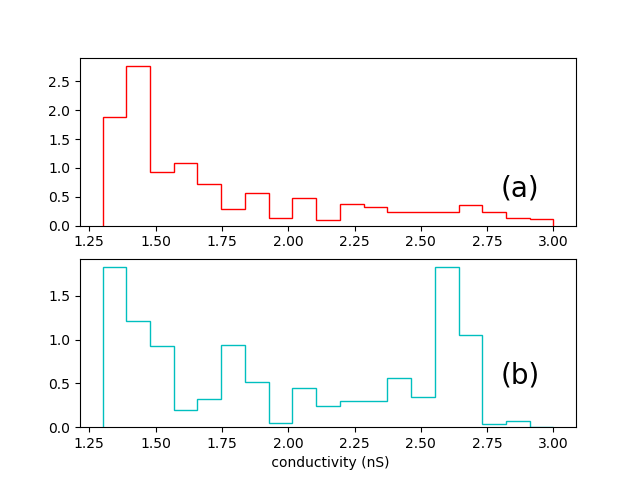}
\caption{ Conductivity histogram (a)[mi22] = 0nM (b) [mi22] = 50nM.}
\label{fig_histo}
\end{figure}

A possible interpretation of the current recordings is as follows.  Each time a nanopore inserts into the bilayer, its resistance decreases by a fixed
amount which depends mainly on the geometry of the nanopore. A simple estimate of the nanopore's conductivity in its closed state 
using a geometrical model which ignores
possible interactions between cations and nanopore's interior yields 1.3nS. In the absence of miR-21, we found conductivity 
distribution centered around this value with a secondary peak  close to 1.6nS, as
shown in the histogram of Fig. \ref{fig_histo}. Previous reports of similar structures  \cite{Howopenclose} also yield values close to 1.6nS. As proven
by numerical simulations \cite{aksi}, transport across DNA nanopores is not only through its lumen, cations can flow along the nanopore's 
outer surface or {\em through the 'gaps' in the DNA structure} \cite{aksi}. This would explain the fact that experimental values can be
larger than theoretical ones. When the nanopore was incubated with the input signal, the 
stem-loop changed conformation as explained above and demonstrated by the coarse-grained simulations. Experimentally, this translated to the
appearance of a second maximum in the distribution of conductivities. Its value ($2.8 \pm 0.2$ nS) is less than twice the value of the 
closed state. This should be expected, as the open stem-loop pushes apart the two halves and at the same time hinders the entrance of the nanopore. 
 
From the present experiments, it is difficult to elucidate further the insertion mechanism of nanopores. A possible interpretation of the
existence of
 transient states could be as follows:
 nanopores lay on one side of the bilayer inserting roughly half of the cholesterol modified strands. This metastable state has
an  energetic penalty due to the exposure of cholesterol to  water. An alternative metastable state corresponds to
a completely inserted nanopore where all the cholesterol moities are in contact with the bilayer's interior and, at the same time, the hydrophylic
outer surface of the nanopore is also in contact with it unless a toroidal rearrangement of lipid heads (as sketched in Fig. \ref{fig_schema}) 
takes place.
 Interaction with streptavidin probably lowers the energetic barrier  between these two metastable states, which would explain the fast dynamics
observed in many recordings. Slow insertion rates could then correspond to insertion in the absence of streptavidin.

\section{Conclusions}
Sensing of short oligonucleotide sequences is potentially an important step in the early detection of diseases such as cancer. 
Developing portable, direct methods to perform such detection could considerably generalize the use of  microARN biomarkers.
Compared to other single molecule detection procedures, nanopore based detection can benefit
from miniaturization techniques used in semiconductor technology, which should provide eventually a
compact, easy to use apparatus. 
In this report, we characterized a DNA nanopore structure which we showed was able to change conformation upon binding with
a DNA analogue of the miR-21 microARN. The conformational change could be characterized by fluorescence and electric recordings.
In doing so, we have shown that detection of single microARNs is a doable task when using the DIB configuration to generate
stable and reproducible bilayers. The major difficulty which remains to be solved is the low rate of insertion into bilayers.
The possibility to detect low concentrations of miRNA depends on the feasability of long electric recordings: the lower the miRNA concentration,
the lower the number of possible open events. Reliable miRNA concentration measurements will therefore not only require parallel measurements but
also a reasonable success rate in the  detection of nanopores.
This seems to be a major hurdle in the design of DNA based nanopores. A possibility explored by other groups was to increase the number of 
hydrophobic moities attached to the nanopore. This is only possible by embeding the  nanopore structure into a larger platform as was done in \cite{LarSim}.

\section{Materials and methods}
\subsection{Fabrication of DNA nanopores}
 DNA nanopores were fabricated in a one-pot
reaction by stepwise cooling an equimolar mixture of staples (1 $\mu$M) in
folding buffer (Tris-acetate-EDTA buffer, 20 mM MgCl$_2$ ) from 85 to 20 $^{\circ}$C in 3h. 
Staple strands were designed using the scaDNAno software.  Before running DIB experiments,
DNA nanopores were further diluted in a 1M KCl buffer containing 0.05\% OPOE (Sigma).
Cholesteryl functionalized DNA pores  were produced by incubating the fully folded pores with cholesteryl
modified strands (Eurogentec) for 45 min with 5 times excess. Before incubation, cholesteryl-modified
oligonucleotides were heated to 60 $^{\circ}$C for 45 min to avoid aggregation.   Streptavidin (Sigma-Aldrich) was
used without any further purification.
\subsection{Lipid preparation}
POPC (1-palmitoyl-2-oleoyl-sn-glycero-3-phosphocholine) and DPhPC (1,2-Dipalmitoyl-sn-glycero-3-phosphocholin)
 were purchased from Avanti Lipids, hexadecane and silicon oil from Sigma. 
Lipids were stocked in chloroform with 10 mg ml$^{-1}$ concentration. Before disolution into a 7:3 hexadecane:silicon oil mixture, 
chloroform was evaporated in a vacuum dissicator for at least  1h. Dissolution of lipids into oil could require a mechanical stirring. 
\subsection{Electric recording}
Two  200pL droplets were deposited in a 60$\mu$L  well machined in poly(methyl
methacrylate)(PMMA). The tip of two silver electrodes  100 $\mu$m in diameter were chlorinated overnight, then
coated with agarose (2\%). The agarose coating facilitated the insertion of electrodes inside the droplets. 
Electrodes were actuated through micromanipulators and connected to an electronic current amplifier (HEKA and Intan). Data were acquired
at 5kHz.

\end{document}